\documentstyle[12pt]{article}
\input epsf
\newlength{\extraspace}
\newlength{\extraspaces}
\newcommand{\be}{\begin{equation}
\addtolength{\abovedisplayskip}{\extraspaces}
\addtolength{\belowdisplayskip}{\extraspaces}
\addtolength{\abovedisplayshortskip}{\extraspace}
\addtolength{\belowdisplayshortskip}{\extraspace}}
\newcommand{\ee}{\end{equation}}
\newcommand{\figuur}[3]{
\hspace{-3mm}{}\ 
\begin{figure}[ht]\begin{center}
\leavevmode\hbox{\epsfxsize=#2 \epsffile{#1.eps}}\\[3mm]
\parbox{14cm}{\small \bf Fig.\ 
\it #3}
\end{center} \end{figure}\hspace{-1.5mm}}

\newcommand{\ba}{\begin{eqnarray}

\addtolength{\abovedisplayskip}{\extraspaces}
\addtolength{\belowdisplayskip}{\extraspaces}
\addtolength{\abovedisplayshortskip}{\extraspace}
\addtolength{\belowdisplayshortskip}{\extraspace}}
\newcommand{\ea}{\end{eqnarray}}

\newcommand{\ppar}{{{}_{\!\!{}/\!/}}}

\begin{document}
\addtolength{\baselineskip}{.2mm}
\renewcommand{\thesubsection}{\arabic{subsection}}

\def\gsim{\mathrel{\raise.3ex\hbox{$>$}\mkern-14mu
             \lower0.6ex\hbox{$\sim$}}}

\begin{titlepage}
\begin{center}

{\hbox to\hsize{
\hfill PUPT-1872}}

{\hbox to\hsize{
\hfill ITFA-99-14}}

\bigskip

\vspace{6\baselineskip}

{\large \sc Holography and Compactification\\}

\bigskip

\bigskip

\bigskip

\bigskip
\bigskip

{\sc  Herman Verlinde}\\[1cm]

{ \it Joseph Henry Laboratories,\\
Princeton University,
Princeton, NJ 08544}\\[.4cm]

{and}\\[.4cm]

{\it Institute for Theoretical Physics,\\ University of Amsterdam}\\
{\it Valckenierstraat 65, 1018 XE Amsterdam}

\vspace*{1.5cm}

{\bf Abstract}\\

\end{center}
\noindent
Following a recent suggestion by Randall and Sundrum, we consider
string compactification scenarios in which a compact slice of
AdS-space arises as a subspace of the compactification manifold.  A
specific example is provided by the type II orientifold equivalent to
type I theory on (orbifolds of) $T^6$, upon taking into account the
gravitational backreaction of the D3-branes localized inside the
$T^6$.  The conformal factor of the four-dimensional metric depends
exponentially on one of the compact directions, which, via the
holographic correspondence, becomes identified with the
renormalization group scale in the uncompactified world. This set-up
can be viewed as a generalization of the AdS/CFT correspondence to
boundary theories that include gravitational dynamics. A striking
consequence is that, in this scenario, the fundamental Planck size
string and the large N QCD string appear as (two different
wavefunctions of) one and the same object.

\end{titlepage}

\newpage

\subsection{Introduction}

In string theory, when considered as a framework for unifying gravity
and quantum mechanics, the fundamental strings are naturally thought
of as Planck size objects. At much lower energies, such as the typical
weak or strong interaction scales, the strings have lost all their
internal structure and behave just as ordinary point-particles. The
physics in this regime is therefore accurately described in terms of
ordinary local quantum field theory, decoupled from the planckian
realm of all string and quantum gravitational physics. The existence
of this large separation of scales has often been used as an argument
against the potential predictive power or even the potential reality
of string theory. However, as we will argue in this letter, there
are a number of recent insights and ideas that suggest a rather more
optimistic scenario, in which string physics may have visible
consequences at much lower energies scales than assumed thus far.  The
two particular recent developments that we wish to combine are:\\

\noindent
(i) {\sc The AdS/CFT correspondence}

The basic example here is the proposed duality between four-dimensional
${\cal N}$ =4 SYM theory and type II B string theory on $AdS_{5} \times
S^{5}$ geometry
\begin{equation}
ds^{2}\,\ =\,\ e^{-2y/R}\,\ {\bf {\it ds}}_{4}^{2}\,\ +\,\ dy^{2}\,\
+R^{2}d\Omega _{5}^{2}  \label{adayes}
\end{equation}
Here ${\bf {\it ds}}_4^2 = dx_\ppar^2$ denotes the flat four-dimensional 
metric and
\begin{equation}
\label{radius}
R^{2}=\alpha'\sqrt{4\pi N g_{s}}\ 
\end{equation}
with $g_{s}= g_{ym}^2$ the type IIB string coupling \cite{jm} 
\cite{jm2}. In this AdS/CFT dictionary, the 
coordinate $y$ needs to be thought of as parametrizing the four-dimensional 
scale: two SYM excitations related by a scale transformation 
\begin{equation}
x_{{\!\!{}_{/\!/}}}\rightarrow e^{\lambda }x_{{\!\!{}_{/\!/}}}
\label{scaletr}
\end{equation}
translate on the AdS-side into two excitations concentrated around 
different locations in the $y$-direction related by a translation 
\begin{equation}
y\rightarrow y+\lambda R.  \label{transl}
\end{equation}
This map thus provides a ``holographic'' projection of the physics of
the gauge theory (which can be thought of as living on the
AdS-boundary) onto the one higher-dimensional AdS space.

Truncating the AdS theory to $y$-values larger (or smaller) than some
finite value $y=y_{0}$ amounts to introducing an UV (or IR) cut-off in
the gauge theory \cite{susswit}. 
In the strict continuum limit, however, the range of
$y$-values extends over the full real axis.  A consequence of this is
that, while the type II theory on the AdS-space contains gravity, the
dual conformal gauge theory on the boundary does not.  Modes of the
AdS gravitational field that extend all the way to the boundary
$y\rightarrow-\infty$ are not normalizable, and therefore do not
fluctuate.

The metric $ds^{2}$ has the rather striking property that for
sufficiently large $y$-values it differs from the four-dimensional
brane-metric ${\bf {\it ds}}_{4}^{2}$ by an exponentially large
``red-shift factor'' $e^{-2y/R}$ , that is, distances along the
$x_{{\!\!{}_{/\!/}}}$-directions are measured very differently by both
metrics. A most dramatic consequence of this that the fundamental IIB
strings, while still of Planck size when viewed by the full metric
$ds^{2}$, can become arbitrarily large when measured in terms
world-brane metric ${\bf {\it ds}}_{4}^{2}$, simply by moving to
larger $y$ values. The SYM interpretation of these large IIB strings
is that they represent the color-electric flux lines \cite{jm2}. 

In principle this type of holographic correspondence can be
generalized to less symmetric gauge theories with non-zero
beta-functions and more complicated phase diagrams. The idea remains
the same, namely that $y$-translations in effect amount to
renormalization group transformations on the gauge theory
side. Beta-functions or symmetry breaking or other type of phase
transitions thus translate into a non-trivial $y$-dependence (e.g. in
the form of domain wall structures) of the metric, dilaton and other
fields relative to the $AdS_5$-background. \\

\noindent
(ii) {\sc Compactifications with a D3-brane world}

In a large class of type I string compactifications, the gauge theory
degrees of freedom in our four-dimensional world may be thought of as
bound to a (collection of) D3-brane(s). These 3-branes wrap our world
but are otherwise localized as point-like objects somewhere inside the
compactification manifold. Scenarios of this type have recently been
studied from various points of view, as in particular they seem to
offer some promising new avenues for addressing the gauge hierarchy
problem \cite{ahdd,rs}. Besides (perhaps) allowing for the possibility
of large extra dimensions \cite{ahdd}, a second new aspect of these
type of compactifications is that (due to the backreaction of the
D3-branes) various fields, such as the dilaton and in particular the
conformal factor of the four-dimensional metric, may acquire
non-trivial dependence on the compact coordinates.

An interesting example of this type of geometry was recently
considered by Randall and Sundrum in \cite{rs}.  In this set-up, the
four-dimensional conformal factor is found to depend exponentially on
the extra fifth coordinate $y$, precisely as in the AdS-geometry
(\ref{adayes}). A new ingredient, relative to the standard AdS
situation, is that in \cite{rs} the extra coordinate $y$ is chosen to
run over a finite (or semi-infinite) range. As a consequence, there
exists a normalizable gravitational collective mode, given by those
fluctuations of the metric $ds^2$ that preserve the form
(\ref{adayes}) but with ${\bf {\it ds}}_4^2$ replaced by a
(sufficiently slowly varying but otherwise) general four-dimensional
metric
\be
{\bf {\it ds}}_{4}^{2}={\bf {\it g}}_{\mu \nu }^{(4)}(x_\ppar)
dx_{{\!\!{}_{/\!/}}}^{\mu
}dx_{{\!\!{}_{/\!/}}}^{\nu }.  \label{fourdee}
\end{equation}
Hence this collective mode behaves just as the ordinary graviton
of our four-dimensional world. In addition, due to the exponential
$y$-behavior of the graviton wave-function, a given object with
five-dimensional mass $m$ has an effective four-dimensional mass
$m_{{4}}$, that depends of its $y$-location via
\begin{equation}
\label{red}
m_{{4}}(y) = m e^{-y/R}
\end{equation}
In \cite{rs} the presence of this exponential red-shift factor (or
``warp factor'') was argued to provide a natural explanation of the
large mass hierarchies, such as between the weak and Planck scale.  \\

\noindent
{\sc Putting} (i) {\sc and} (ii) {\sc together}:\\
It seems natural to look for a way of combining these two ideas.
Concretely, one could ask the following two (probably equivalent)
questions:

\begin{itemize}
\item{Can the AdS/CFT type of dualities be extended to situations where 
the radial AdS-coordinate is effectively compactified, and does this 
compactifiation indeed automatically imply the presence of four-dimensional 
gravity on the boundary?}
\item{Are there string compactification scenarios where a
compact slice of AdS-space appears as part of the compactification
geometry, and does the radial AdS coordinate then again
have a ``holographic'' interpretation as parametrizing the RG scale
of the four-dimensional theory?}
\end{itemize}

\noindent
In the following we will argue that both questions have a positive
answer, though the first one only for a (large but) finite set of
D3-brane charges $N$. Consequently, the compactified AdS-geometry will
be a good approximation provided the string coupling $g_s$ is not too
small compared to $1/N$.  In the concluding section we will address
some of the possible consequences of this new view on string
compactification.

\figuur{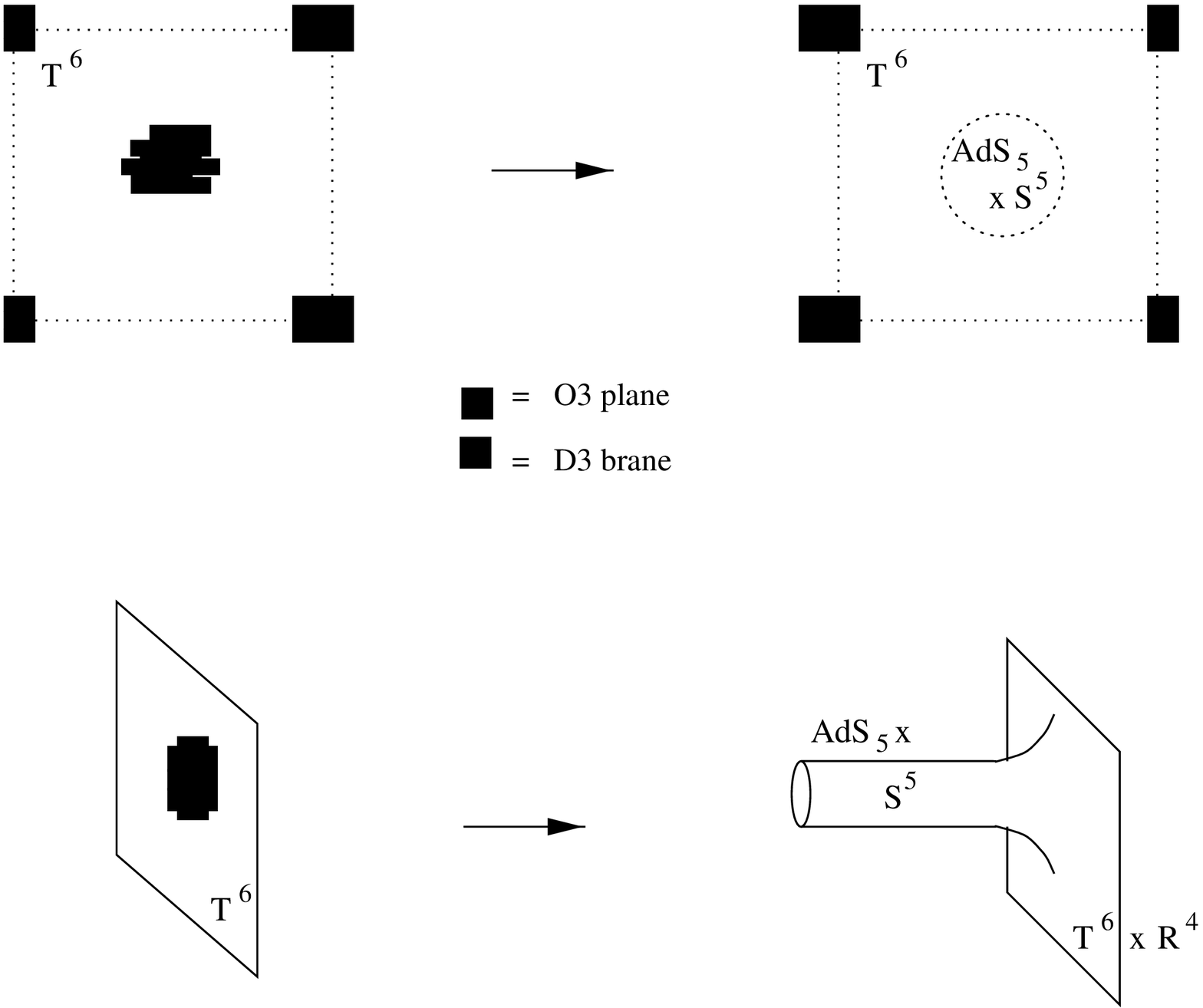}{11cm}{At the $U(N)$ symmetric point,
when $N$ D3-branes coincide, their backreaction produces a (compact)
$AdS_5 \times S^5$ sub-region, that can be glued into the $R^4 \times
T^6$ geometry.}

\subsection{AdS Compactification}

Let us start with considering the toy example 
of a four-dimensional ${\cal N}$=4 supersymmetric world described
by the $T^6$ compactification of the type I superstring. Its low energy 
effective description is given by ${\cal N}$=4 super-Yang-Mills theory, 
with as maximal unbroken gauge symmetry group ${\cal G} = SO(32)$, 
coupled to supergravity. In the following we will consider the 
situation where this symmetry is broken down to an $U(N)$ sub-group 
with $N\leq 16$.  

By applying $T$-duality this type I string theory can be equivalently
described as an orientifold of type II string theory on $T^6$. In this
representation, there are $2^6 = 64$ orientifold planes located at all
the half-way points of the $T^6$. In addition, there are 32 D3-branes
inside the $T^6$ which are pairwise identified under the orientifold
${\bf Z}_2$-action \cite{jtasi}.  Hence it is possible for $N\leq16$
of these D3-branes to form a small cluster inside the $T^6$, and in
the limit where all $N$ coincide at the same point, the unbroken
$U(N)$ gauge symmetry appears.

At low energies in the uncompactified world, gravity effectively
decouples from the $U(N)$ SYM dynamics. Thus we can look for a
regime of parameters in which we can apply the AdS/CFT duality map
and obtain type II theory on $AdS_5 \times S^5$ as a good dual 
low energy description. The main restriction (in order to be 
able to trust the sigma-model approximation) is that the $AdS_5$
radius of curvature $R$, given in (\ref{radius}), needs to be
large compared to the string scale, 
\be
\label{ineq}
g_s >\!\! > {1\over 4\pi N}.
\ee
Hence, relative to the usual large $N$ context, we now have a somewhat 
more limited control over the approximations involved in the duality 
map.

Around the string scale $L_S$, open string effects will start to
modify the SYM dynamics. Let us choose the radial AdS-coordinate $y$
such that this string scale $L_S$ gets mapped to the region around
$y\simeq 0$. Then the $AdS_{5} \times S^5$-description is a valid
approximation for the low-energy regime $y>\!\! > R$.  Around $y\simeq 
0$, on the other hand, we reach the transition region where the low 
energy regime meets with the high energy regime described by orientifold of
type II on $T^6$. Our main new proposal is that these two regimes can
be consistently glued together into one single dual type II string
background.  In the following we will assume that the size of the 
$T^6$ is bigger or of the order of the AdS-radius $R$ in (\ref{radius}).

As a first approximation, we can visualize the total target space of
this new type II theory as follows. We first cut out a small ball
around the D3-branes inside the $T^6$, such that its outside radius
coincides with the $S^5$ radius $R$ in (\ref{radius}). Then we replace
the inside of the ball by the $y\gsim 0$ region of $AdS_5 \times
S^5$. (See fig. 1).  It seems reasonable to assume that the resulting
total space can be smoothed out to obtain an exact consistent type II
background, and the end result of this procedure should be equivalent to
taking into account the gravitational backreaction of the D3-branes.

In a more complete treatment, we must also include the backreaction
of the 64 orientifold planes. These have a negative tension equal to
$-1/4$ times the D3-brane tension, which need to be taken into account. 
we can write an explicit form for the background metric, by 
starting from the general form
\be
\label{hades}
ds^2 = {1\over H(x_\perp)^{1/2}} \, {\bf \it ds}_4^2 \, + \,
H(x_\perp)^{1/2} \, dx_\perp^2.  
\ee 
with ${\bf \it ds}_4^2 =
dx_\ppar^2$, and where $x_\perp$ denote the coordinates inside the
$T^6$. For simplicity, let us consider the most symmetric example 
of the $SO(32)$ invariant point, where
all 16 D3-branes coincide with their 16 ${\bf Z}_2$ images at one
of the orientifold fixed points, say at $x_\perp \! = 0$. 
Taking the $T^6$ to be an exact cube with period $R_c$, the harmonic
function $H(x_\perp)$ then takes the form
\be
H(x_\perp) = 1 + 4 \pi g_s (\alpha')^2 \Bigl[ f_D(x_\perp) - 
f_O(x_\perp) \Bigr]
\ee 
where 
\be
f_D(x_\perp) = 2 \sum_{\vec{n} \in {\bf Z}^6} \; 
{16\over |\vec{x}_\perp +\vec{n}{R}_c|^4} \, 
\ee
denotes the contribution of the 16 D3-branes and their ${\bf Z}_2$-image, and
\be
f_O = 
2 \sum_{\vec{n} \in {\bf Z}^6}  {1/4 \over |\vec{x}_\perp 
+ {1\over 2}\vec{n} R_c|^4}
\ee
that of the 64 orientifold planes located at all the half-way points
inside the $T^6$.  In the region close to the D3-branes at $x_\perp = 0$, 
the above geometry indeed reduces to the $AdS_5 \times S^5$ metric
(\ref{adayes}).\footnote{For the $SO(32)$ symmetric example 
at hand, the $S^5$ must in fact be replaced by its orientifold
$S^5/{\bf Z}_2$.  We further notice that close to the other
orientifold planes the function $H$ goes through 0; in this horizon
region, however, $\alpha'$ and non-perturbative string-corrections are
expected to play an important role.}

The present minimal example (of the simple $T^6$ orientifold
compactification) can straightforwardly be generalized to less
symmetric theories obtained e.g. by applying further orbifoldings.  In
this way one can find holographic duals to a large class of
four-dimensional type I string theories, typically with a reduced
number of (or even zero) supersymmetries.  The total target space of
the type II duals should all describe consistent compactifications of
the $AdS_5\times S^5$ geometry. The number of such consistent
AdS-compactifications, however, is constrained by the usual tadpole
cancelation conditions. In particular, one needs to make sure that the
RR flux $N$ that escapes the AdS-region gets absorbed by the
appropriate number of branes and orientifold planes present in the
compact outside region. Consistent solutions to this condition can be
found only for a finite set of $N$-values. Still, relative to our
minimal set-up, one can significantly extend the range of allowed
values for $N$, up to values of the order of $N \simeq 10^3$ or
larger, by considering more general orbifold spaces. Hence in this
more general setting, the restriction (\ref{ineq}) in effect just
amounts to $g_s >\!\! > 10^{-4}$, which leaves a quite sufficient
parameter range for which we can trust our approximations.

\subsection{Gravitational coupling}

This geometric set-up is now indeed very similar to that of \cite{rs},
where our $T^6$ region with the orientifold planes plays the role as
the positive tension brane in their scenario.  Several of the same
conclusions apply. In particular, we must find a collective graviton
mode of the exact same type as described in the Introduction, that
couples precisely as the four-dimensional graviton to the boundary
field theory. In our case this result was of course expected from the
start, since the boundary theory is in fact the type I string theory
on $T^6 \times {\bf R}^4$, that also contains a gravitational closed
string sector.

The relation between the four-dimensional Planck length
$L_{pl}$ and the 10-dimensional one $L_{\! {}_{10}}$ takes the
usual form \be (L_{\! {}_{10}})^8 = (L_{pl})^2 \, V_6 \ee where $V_6$ is the
appropriately measured volume of the $T^6$.  Plugging the form
(\ref{hades}) into the ten-dimensional action gives 
\be 
{1\over (L_{\!
{}_{10}})^8 }\int \! d^{10} x \,   {\textstyle \sqrt{-g}}{}_{\! {}_{10}} \, 
 R_{\! {}_{10}}
= {V_6 \over (L_{\! {}_{10}})^8} \int \! d^{4} x_\ppar \, 
 {\textstyle \sqrt{-g}}{}_{{4}} R_{{4}} 
\ee 
with 
\be 
V_6 = \int_{\strut T^6} \! d^6
x_\perp \, H(x_\perp).  
\ee 
Notice that, in spite of the fourth order
pole in $H(x_\perp)$ near the branes, this integral indeed yields a
finite answer.

The collective mode of the 10-dimensional metric that produces the
four-dimensional graviton again has the same shape as the full metric,
as given in eqn (\ref{hades}), with $ds_{4}^2$
a general four-metric (though sufficiently slowly varying with
$x_\ppar$).  The shape of the graviton wave-function the AdS-region
decays with $y$ as $e^{-2y/R}$. This exponential factor is reflects
the fact that the coupling of matter to the graviton becomes weaker
and weaker at lower and lower scales. 
Indeed, we can extract the strength of the coupling to matter by inserting 
the variation $\delta g^{(4)}_{\mu\nu} = h_{\mu\nu}$ of the four-metric 
$ds_4^2$ into the the ten-metric (\ref{hades}). The corresponding
variation of the 10-d matter action reads
\be
\int \! d^{10} x \,  {\textstyle \sqrt{-g}}{}_{\! {}_{10}} \, 
\delta g_{\! {}_{10}}^{\mu\nu} \,
T_{\mu\nu} \,
= \, \int \! d^4 x_\ppar \, h^{\mu\nu} \, \overline{T}_{\mu\nu}
\ee
with 
\be
\overline{T}_{\mu\nu}(x_\ppar) = 
\int \! d^6 x_\perp \, H(x_\perp)\, T_{\mu\nu}.
\ee
Inserting on the right-hand side the energy-momentum tensor 
(here all indices contracted with the ten-metric $ds^2$)
\be
T_{\mu\nu} = {m \over \sqrt{- g_{\! {}_{10}}}}\, \int \! d\tau\;\, 
{\dot{x}_\mu\dot{x}_\nu \over \sqrt { \dot{x}^2}} \;\, 
\delta_{{}_{10}}(x \! - \!
x(\tau))
\ee 
of a 10-d point-particle, with mass $m$ located at a given 
location $y_\perp$, gives for the effective 4-d energy-momentum
tensor (with all indices contracted with the four-metric $ds_4^2$)
\be
\overline{T}_{\mu\nu} = {m \over  H(x_\perp)^{1/4}} \, \int \! d\tau\;\; 
{\dot{x}_\mu\dot{x}_\nu \over \sqrt { \dot{x}^2}} \;\;
\delta_{{4}}(x_\ppar \! -\! x_\ppar(\tau))
\ee
So the four-dimensional mass depends on $x_\perp$ as
\be
m_{4}(x_\perp) =  {m \over H(x_\perp)^{1/4}}.
\ee
This is the expected red-shift effect; close to the D3 branes it reduces 
to (\ref{red}).

\subsection{RG scale as a real extra dimension}

Though familiar and standard, the appearance of both open and closed
string dynamics in type I string theory is still a deep fact. The two
are indeed intertwined in an intricate way, since when one
starts including the open string quantum effects, there is no
obviously unique way of separating the open string loop diagrams from
closed string tadpoles. This very subtlety of course lies at the heart
of the D-brane correspondence between gauge theory and gravity.  
A microscopic reconstruction of the reasoning of \cite{jm}
indeed shows that the appearance of the curved AdS-metric (\ref{adayes}),
as describing the near-horizon region close to the D3-branes, must be 
thought of as a cumulative consequence of open string quantum effects.

Concretely, one can imagine setting up a renormalization group flow,
where via a Fischler-Susskind type mechanism, the effect of
integrating out successive momentum shells in the open string channel
gets absorbed into an appropriate redefinition of the world-sheet
sigma-model couplings.  Roughly speaking, in this procedure the slice
of the AdS target space in between two values $y_{1}$ and $y_{2}$ gets
created from the open string dynamics in between the corresponding
energy scales. Eventually, after all open strings are integrated out,
its planar diagrams have been replaced by a stretched-out closed
string worldsheet, moving inside the semi-infinite AdS-throat
geometry. Thus, in a quite precise sense, the AdS closed strings
provide the long sought after dual representation of the gauge theory
planar graphs \cite{jm} \cite{jm2}.

If instead one runs this renormalization group flow from the string
scale down to a certain finite length scale $L$ (for example the weak
scale), the resulting target space consists of a compact slice of AdS
bounded by the Planck region $y \simeq 0$ on one side and by an
effective D3-brane located at $y \simeq R \log L/L_{S}$ on the other.
The D3-brane hosts the remaining sector of low energy open strings\footnote{Due to the redshift, these open strings have acquired a
renormalized tension $\alpha'_{ef\!f} \simeq \sqrt{Ng_{s}}\; L^{2}$.
The energy of a string with this tension stretched over a distance $L$
reproduces a force comparable to the $1/L$ Coulombic force of the SYM
model.}, describing the physics at distance scales larger than
$L$.  Sub-planckian physics thus happens far inside the AdS-tube,
and from this perspective the $T^6$-orientifold region merely acts as
a kind of sounding board, providing some appropriate boundary
condition in the far away planckian regime.  Via high energy
experiments on the D3-brane world-volume, however, one can still
generate quantum fluctuations that extend to smaller values of $y$ and
thereby probe the corresponding planckian geometry.

\medskip

What new lesson can we extract from this? It is clear that, relative
to the standard notion of an extra dimension, the $y$-direction is
rather unusual. Normally one would expect that space-like separated
events are independent, while here we are learning that
$y$-translations are in effect scale transformations. Therefore, to
avoid over-counting of the number of degrees of freedom, we thus
indeed need to
adopt the hypothesis that physics that happens far out in the
$y$-direction is not really independent from that, say, at the
boundary region $y \simeq 0$, but rather a ``holographic image'' of
physics at $y\simeq 0$ happening at a corresponding scale $L(y)= L_S
e^{y/R}$.\footnote{Because of this holographic identification, the
distinction made in \cite{rs} between a ``hidden'' and ``visible''
brane is no longer applicable in our context.  Also the (continuum) KK
tower of gravitational excitations must be reinterpreted as actually
representing low energy SYM degrees freedom in the boundary theory.}
This is the familiar dictionary of the AdS/CFT
correspondence \cite{gkp}\cite{wads}.

Though quite miraculous, in itself this holographic equivalence
doesn't teach us anything new yet, as it simply points to some
redundance in the description. We could for example insist on
describing all the physics as happening just at $y\simeq 0$, and
consider the holographic map as some purely mathematical
equivalence. Relative to the standard AdS/CFT situation, however, we
now have four-dimensional gravity as an essential new ingredient. In
first approximation the gravity coupling seems quite consistent with
holography, since (as seen from the previous section) it looks like to
a relatively standard Kaluza-Klein dimensional reduction. The
holographic reconstruction does, however, strictly amount to a
non-trivial redistribution of matter, and since gravity is still in
the game, in the end we do need to specify which energy-momentum
distribution acts as a source for the gravitational field.

The new consequence of this holographic view of gravity, therefore, is
that the renormalization group scale is promoted to a real physical
extra direction. Concretely, this means that for type I string theory
the gravitational energy-momentum in our four-dimensional world really
spreads out into the extra $y$-direction, indeed similar to a
holographic, or in a perhaps more accurate analogy, chromatographic
image.

\subsection{Concluding remarks}

Finally, let us try to draw a few obvious and less obvious conclusions. 
We begin with \\

\noindent
{\it The Planck vs large N QCD string}

A first striking property of this new scenario is that the planckian
type II string and the low energy electric flux strings are in essence
made up from one and the same object: the two are simply related by a
translation in the $y$-direction. Similarly, in more realistic type I
compactifications with QCD-like confining gauge sectors, one expects
to find glueball excitations that on the type II side correspond to
bound state solutions to the graviton wave-equation, localized
(possibly around some domain wall structure) far inside the
AdS-region. This is just like in the standard AdS/CFT correspondence.
However, the new ingredient here is that the bulk gravity (used for
representing the glueballs) and the gravity in the physical ``boundary
world'' are directly linked and produced by the same type II string.\\

\noindent 
{\it Low energy string phenomenology}

The above set-up seems to open up the possibility of finding type II
string compactification scenarios that in effect bypass the quark and
gluon stage and immediately connect with the infra-red degrees of
freedom in terms of mesons, baryons, etc. More generally, instead of
the usual ``top-down'' approach to string phenomenology, one may
contemplate a rather different ``bottom-up'' philosophy.  Ideally, one
could first try to connect string theory with low-energy physics
(e.g. via standard non-compact AdS/CFT type technology and variations
thereof) and then afterwards introduce gravity by trying to look for
all possible consistent compactifications of this AdS-type space. Of
course, there are still important technical as well as conceptual
obstacles to deal with, such as supersymmetry breaking, unification,
massless moduli fields, and allowing $N$ to be small rather than
large, just to name a few. \\

\noindent 
{\it The gauge hierarchy problem}

Perhaps the deepest consequence of the above picture is that the
renormalization group scale parameter of the four-dimensional world
gets promoted to a real physical extra dimension. In this way all
relative mass hierarchies, such as those between the weak scale and
the Planck scale, are translated into relative separations in this
extra direction. Since the conformal factor decays exponentially with
$y$ with a planckian decay length, separations of say 50 or 100 times
the Planck length generate scale factors as large as $10^{15}$ or
$10^{30}$ \cite{rs}.  While perhaps one could argue that this gives a
natural explanation of the large mass separations in our world,
without giving a good reason of why string theory would necessarily
choose this type of compactification geometry, one cannot claim it
really solves the hierarchy problem. Instead, it seems more sensible
to turn the conclusion around, and consider the existence of a large
mass hierarchy as evidence supporting this type of compactification
scenario as its most natural geometric realization.\\

\noindent 
{\it The cosmological constant problem}

The most vexing hierarchy problem is of course that of the
cosmological constant. In essence it arises as a clash between
experiment and the theoretical knowledge of general relativity and the
renormalization group. There are several ingredients in the present
set-up, however, such as (i) the reinterpretation of the RG scale as a
holographic extra dimension, (ii) the intimate connection between the
RG flow and the AdS-gravity equations of motion, (iii) the presence of
a planckian size negative cosmological constant inside the AdS-region,
that suggest a rather new perspective on this problem.

In this connection, it seems interesting to note that a naive AdS
holographic description of the universe (as a boundary theory with
finite temperature equal to that of the microwave background)
identifies the total geometry with that of an AdS black hole
\cite{ewtemp} with a horizon radius of about a millimeter, and thus
with associated entropy of the order of $10^{90}$. A suggestive
coincidence is that this horizon size is of similar magnitude as the
thermal wavelength of the radiation itself, which perhaps indicates
that holography may provide a link between the size and the total
entropy of the universe \cite{erik}. 

Finally, we notice that this view of the universe as an AdS black hole
has the striking consequence that any pair of points 
can be connected via a path of at most a couple of millimeter
length. Indeed, one can simply first travel close to the black hole
horizon, traverse the required distance in the $x_\ppar$-direction,
and then go back. This does not mean, however, that one can travel at
super-luminous speeds, since the travel time (as measured in our
world) is delayed correspondingly by the red-shift factor.\\

\bigskip

\noindent 
{\sc Acknowledgements}

This work is supported by NSF-grant 98-02484, a Pionier fellowship of
NWO, and the Packard foundation. I would like to thank Costas Bachas,
Micha Berkooz, Jan de Boer, Robbert Dijkgraaf, Savas Dimopoulos, Lisa
Randall, Savdeep Sethi, Raman Sundrum, Erik Verlinde, and E. Witten 
for very helpful discussions.\\

\bigskip

\bigskip
\renewcommand{\Large}{\large}

\end{document}